Research Papers

ECIS 2018 Proceedings

11-28-2018

# TOWARDS A KNOWLEDGE LEAKAGE MITIGATION FRAMEWORK FOR MOBILE DEVICES IN KNOWLEDGE-INTENSIVE ORGANIZATIONS

Carlos Andres Agudelo-Serna
*The University of Melbourne*, cagudelo@student.unimelb.edu.au

R Bosua
*The University of Melbourne*, rachelle.bosua@ou.nl

Atif Ahmad
*The University of Melbourne*, atiff.ahmad@unimelb.edu.au

Sean B. Maynard
*The University of Melbourne*, seanbmaynard@unimelb.edu.au

Follow this and additional works at: https://aisel.aisnet.org/ecis2018_rp

# TOWARDS A KNOWLEDGE LEAKAGE MITIGATION FRAMEWORK FOR MOBILE DEVICES IN KNOWLEDGE-INTENSIVE ORGANIZATIONS

*Research paper*

Agudelo-Serna, Carlos, University of Melbourne, Australia, cagudelo@student.unimelb.edu.au

Bosua, Rachelle, University of Melbourne, Australia, rachelle.bosua@unimelb.edu.au

Ahmad, Atif, University of Melbourne, Australia, atif@unimelb.edu.au

Maynard, Sean B., University of Melbourne, Australia, seanbm@unimelb.edu.au

## Abstract

*The use of mobile devices in knowledge-intensive organizations while effective and cost-efficient also pose a challenging management problem. Often employees whether deliberately or inadvertently are the cause of knowledge leakage in organizations and the use of mobile devices further exacerbates it. This problem is the result of overly focusing on technical controls while neglecting human factors. Knowledge leakage is a multidimensional problem, and in this paper, we highlight the different dimensions that constitute it. In this study, our contributions are threefold. First, we study knowledge leakage risk (KLR) within the context of mobile devices in knowledge-intensive organizations in Australia. Second, we present a conceptual framework to explain and categorize the mitigation strategies to combat KLR through the use of mobile devices grounded in the literature. And third, we apply the framework to the findings from interviews with security and knowledge managers.*

# 1    Introduction

Knowledge and information leakage represent one of the most common security risks faced by organizations worldwide. Recent studies have shown how organizations struggle with leakage of sensitive organizational information across various avenues, such as social media, cloud computing and portable data devices (Agudelo-Serna et al., 2017; Ahmad et al., 2014, 2015; Jiang et al., 2013; Krishnamurthy and Wills, 2010; Mohamed et al., 2006; Ponemon Institute, 2016). Although much of the literature has focused on technical security-related aspects of leakage (i.e., data and information), limited research has been conducted on knowledge leakage through mobile devices (Agudelo et al., 2015; Ghosh and Rai, 2013; Zahadat et al., 2015).

Although the use of mobile devices (whether employee or organization owned) has shown to be convenient in the context of higher mobility, such apparent benefit comes at a high security cost. By using these devices in knowledge-sharing activities, knowledge workers expose themselves to confidentiality risks. Often these challenges in confidentiality occur as a result of employees' security (mis)behaviors.

The abundant literature on leakage in terms of data and information (Abdul Molok et al., 2010a; Chen et al., 2014; D'Arcy et al., 2009a; Gordon, 2007; Krishnamurthy and Wills, 2010; Morrow, 2012; Shabtai et al., 2012; Yahav et al., 2014) markedly contrasts with that of knowledge leakage research, which comparatively, continues to be underrepresented in the current knowledge management literature. Nevertheless, most recent literature has emphasized the criticality of knowledge protection in organizations as a way to sustain competitive advantage and the importance of the human dimension whilst dealing with knowledge assets (Ahmad et al., 2014, 2015; Durst et al., 2015; Jiang et al., 2016; Kang and Lee, 2017; Mupepi, Mambo Governor Modak et al., 2017; Sumbal et al., 2017; Tan et al., 2016; Tsang et al., 2016; Zhang et al., 2017).

Thus, we contend that knowledge leakage is a multi-dimensional problem and subsequently the focus should not only be on technological (e.g., firewall, antivirus, and compartmentalization) and formal (i.e., policies, standards and procedures) controls, but also on human factors and informal controls as well (Agudelo et al., 2015, 2016; Ahmad et al., 2014; D'Arcy et al., 2009a; Jiang et al., 2016; Mupepi, Mambo Governor Modak et al., 2017).

Often, workers, rather than hackers, are usually the main sources of information breaches whether deliberately or inadvertently. As a matter of fact, research has shown that the culture and people within an organization are just as likely to be the source of data leakage (Abdul Molok et al., 2011a; Ahmad et al., 2015; Colwill, 2009; Crossler et al., 2013). For example, confidential and sensitive information is sometimes shared inadvertently through social media and mobile devices. Research shows that despite the security policies, procedures, and tools currently in place, employees around the world engage in risky behaviours that risk corporate and personal data. Employee behaviours include unauthorised application use, misuse of corporate computers, unauthorised physical and network access, remote worker security, misuse of passwords, amongst others (Abdul Molok et al., 2011b; Agudelo-serna et al., 2017; Agudelo et al., 2016; Yahav et al., 2014).

To reduce the leakage risk and protect corporate knowledge, businesses must integrate security into the corporate culture and consistently evaluate the risks of every interaction with networks, devices, applications, data, and other users. Organizations need to understand how employee behaviour increases risk and take steps to foster a security-conscious corporate culture in which employees adhere to policies and procedures. In other cases, addressing human aspects through deterrence with education and awareness programs have proved to be effective (D'Arcy et al., 2009a; Tsang et al., 2016). Additionally, mobility and mobile technology have raised the risk profile of organizations (Derballa and Pousttchi, 2006a, 2006b; Zhang et al., 2017).

In this empirical study, we concentrate on the gap found in the literature and practice. By conducting a series of interviews with information security and knowledge experts in knowledge-intensive organizations in Australia, we attempted to answer the following research question:

*How can knowledge-intensive organizations address the risk of knowledge leakage (KLR) caused by the use of mobile devices?*

As part of a larger research that seeks to understand how organizations protect their knowledge to guard against the erosion of competitive advantage in an increasingly mobile and interconnected society, this paper will focus only on the specific strategies to address KLR caused by the use of mobile devices (and technology) in the context of Australian organizations.

This study takes a contextual approach to understand how KLR changes depending on the circumstances within which knowledge leakage occurs and uses a conceptual model to explain the factors that affect the risk of leakage through the use of mobile devices. This understanding is crucial because by learning the determinants behind this risk, organizations can effectively develop more efficient formal (policy), informal (culture, behavior, Security Education Training and Awareness) and technological safeguards (Ahmad and Maynard, 2014; D'Arcy et al., 2009b; Dhillon, 2007).

The remaining of the paper is structured as follows: First, it provides salient concepts found in the key background literature. Second, the conceptual model is shown followed by a brief discussion of the constructs, third, the initial analysis of the interviews follows. Finally, the potential contributions and future work are presented.

## 2    Related Work

### 2.1    Knowledge

The difference between data, information and knowledge is addressed in multiple sources such as Boisot and Canals (2004) and Dahlbom and Mathiassen (1993). Boisot and Canals (2004) state that raw data becomes information through processes that add meaning from such data, and, adding the contextual understanding in conjunction with the background of such data allows knowledge to be inferred. Therefore, knowledge is intertwined with data and information. Consequently, knowledge leakage is also related to data and information leakage. This distinction is important for our study because the leakage of data/information may also result in knowledge leakage just by drawing on inferences, that is, humans gain knowledge by inferences – the process of inferring things based on what is already known (Dahlbom and Mathiassen, 1993). According to Schwartz (2006) as a way of circumventing to a certain extent this debate about knowledge, information and data (KID) regarding the granularity of knowledge, we should take the perspective of knowledge management (KM) process and focus on praxis, "*taking as a starting point the question, What do we need to do with knowledge in order to make it viable for an organization to use, reuse, and manage it as a tangible resource, and apply it toward specific actions?*" (p. 11). Schwartz (2006) also argues that such KID distinction although important can be conveniently ignored as it is not essential to the fundamental mission of KM. In taking this approach, we acknowledge that knowledge leakage is a KM issue and posit that knowledge should be analyzed from an applied pragmatic and holistic view (Schwartz and Te'eni, 2011).

This study adopts the definition of knowledge given by Davenport and Prusak (1998):

*"Knowledge is a fluid mix of framed experience, values, contextual information, and expert insight that provides a framework for evaluating and incorporating new experiences and information. It originates and is applied in the minds of knowers. In organizations, it often becomes embedded not only in documents or repositories but also in organizational routines, processes, practices, and norms."*

According to this definition, knowledge is complex, a mixture of various elements, intuitive and therefore, hard to capture. Moreover, knowledge is embedded in people, and as such, may be unpredictable and intangible. Knowledge derives from information and to turn information into knowledge, human mediation is required (Davenport and Prusak, 1998). Although knowledge is further divided into tacit (present in employee's minds) and explicit (knowledge that has been codified into artefacts) (Nonaka, 1991), from the perspective of mobile devices, this study will only focus on explicit knowledge leakage, since its disclosure is more likely to occur in mobile device (and mobile technology) settings than tacit knowledge leakage, such as in situations where key personnel leave the organization to a competitor (Frishammar et al., 2015).

Information and knowledge have become key strategic assets (Bollinger and Smith, 2001; Grant, 1996; Spender and Grant, 1996) for knowledge-intensive organizations to achieve sustained competitive advantage, innovation and value creation (Nonaka and Toyama, 2003; Sveiby, 1997). Further, Blackler, (1995) describes the knowledge work undertaken in these organizations as the direct manipulation of symbols to create an original knowledge product, or to add obvious value to an existing one, i.e, creative work. Knowledge work also includes the handling and distribution of information where all workers involved in the chain of producing and distributing knowledge products emphasize non-routine, problem solving approach that requires a combination of convergent and divergent thinking.

Similarly, MacDougall and Hurst (2005) contend that the adoption of knowledge workers, employees who produce value by utilizing their knowledge rather than physical labor, allows organizations to develop their knowledge assets. These individuals perform work based on their information assets for the coordination and management of organizational activities (Sorensen et al., 2008). Ristovska et al. (2012) also focus on the importance of knowledge embedded in knowledge workers as it is an organizational asset for achieving sustainable competitive advantage which can be materialized into documentation and organizational processes. The importance of expertise in organizations relies heavily on exercising specialist knowledge and competencies, or alternatively, the management of organizational competencies and capabilities which belong to employees or knowledge workers (Blackler, 1995; Thompson and Walsham, 2004).

Consequently, knowledge, in this context, is the mix of contextual information existing in the mind of the knowledge worker, tailored by the individual, based on facts, procedures, concepts, interpretations, ideas, observations and judgments which is then codified into artefacts such as processes, guidelines and documentation in organizations (Alavi and Leidner, 2001).

## 2.2    Knowledge leakage

Knowledge leakage (KL) is defined in this paper: as the **accidental** or **deliberate** loss or unauthorized transfer of organizational knowledge intended to stay within a firm's boundary resulting in the deterioration of competitiveness and industrial position of the organization (Frishammar et al., 2015; Nunes et al., 2006).

According to the knowledge leakage definition, KL can occur from the disclosure of sensitive details, information or data as meaning can be inferred by a competitor based on understanding of context and leveraged even further to generate insights and advance their own competitiveness to the detriment of the organization's competitive advantage (Abdul Molok et al., 2010b; Ahmad et al., 2015; Annansingh, 2012; Davenport and Prusak, 1998).

As mentioned above, knowledge leakage, in the meaning of knowledge leaking away from its origin, can occur in different situations. However, recent research has also shown the dichotomy of leakage as it can be considered to be positive, when the organization benefits from it, or negative, when it is detrimental to the organization (Grillitsch and Nilsson, 2017; Jiang et al., 2016).

Therefore knowledge leakage does not necessarily have a negative connotation. For example, in collaboration, a positive knowledge leakage can occur in the form of knowledge spillover between cooperation partners (Mupepi, Mambo Governor Modak et al., 2017). On the other hand, a negative knowledge leakage occurs when an actor, consciously or inadvertently, leaks knowledge about the focal firm to a competitor (Durst et al., 2015).

Despite the criticality that knowledge leakage can have on organizations in either direction and the fact that knowledge management practices, such as knowledge transfer or acquisition have been studied extensively, the study of knowledge leakage appears to be insufficiently researched (Kang and Lee, 2017).

Although knowledge loss due to a lack of knowledge management procedures is also defined as knowledge leakage (Nunes et al., 2006), the focus in this study will be on negative knowledge leakage directly or indirectly caused by knowledge workers when performing knowledge work through **mobile devices**, particularly, the accidental loss derived from misbehaviors (failing to comply policy and procedures), as it is considered the most challenging channel of leakage for organizations to control (Durst et al., 2015; Nunes et al., 2006).

Similarly, the inadvertent loss, caused by insiders can be influenced by addressing human behavior, habits through policy, culture and awareness as opposed to malicious insiders who are deliberately seeking to leak knowledge/information (Colwill, 2009; D'Arcy et al., 2009a) and are not influenced by such controls. Therefore, the focus on this study will be on addressing unintentional negative leakage caused by non-malicious insiders in the context of mobile devices.

Drawing upon the standard definition of risk, knowledge leakage risk (KLR), in this study, is defined as the probability that KL occurs multiplied by the impact of the KL to the organization (Tsang et al., 2016). That is, ***KLR = Probability of Occurrence x Impact of KL***

## 2.3    Mobile Usage Contexts

In this study, the definition of mobile device refers to an autonomous, portable, and wireless computing device. Such a device is characterized by mobility and used to perform knowledge work typically but not exclusively away from the organizational physical boundaries. Examples of mobile devices include smartphones, tablets and laptops. (Agudelo-serna et al., 2017; Mansfield-Devine, 2012; McLellan, 2013; Tu et al., 2015).

To address knowledge leakage risk through mobile devices and mobility in organizations, we take a context-specific approach to recognize how risk changes according to the circumstances and factors within which leakage occurs.

Although knowledge leakage is driven by the worker in control of the mobile device, there are multiple environmental factors that affect the use of mobile devices for knowledge work. Nonaka and Toyama (2003) suggest that knowledge creation, sharing and distribution are achieved through the interactions between the individual, the organization and the environment. The environment influences the individual while, at the same time, individuals are continuously recreating their environment through their social interactions. This proposes that social factors in human interactions constantly change the environment in which knowledge is created. Nonaka and Toyama (2003) developed a model of knowledge creation in order to explain the conversion of knowledge through interactions between individuals, groups of individuals, organizations and the environment. This model not only highlights the importance of the environmental and organizational circumstances around an individual, it also highlights the importance of the social environment where individuals interact within groups to obtain information (Nonaka and Konno, 1998; Nonaka and Toyama, 2003). The concept of mobility in knowledge management is not new and the substantial impact of mobile technology on knowledge management processes has been previously identified leading to the definition of *mobile knowledge management* and the importance of mobile environments (Derballa and Pousttchi, 2006a, 2006b).

These environments are referred to in the literature, from a mobile device perspective, as the "context" of the mobile device usage (Abdoul Aziz Diallo, 2012; Chen and Nath, 2008; Schilit et al., 1994). Table 1 summarizes the mobile-usage-context taxonomy developed from the literature. In understanding the different contexts of mobile-device and mobility in these different settings (technological, environmental, organizational, social and personal), it is important to assess the overall security risk of the device as the potential enabler of, or medium through which knowledge leakage can occur, in conjunction with the user and the environment within which the device is used. The importance of mobile device contexts stems from the fact that without the context within which knowledge leakage occurs, it is not possible to determine the level of risk (Benítez-Guerrero et al., 2012; Bradley and Dunlop, 2005). This concept also highlights the strong relationship between mobile and mobility, while the former refers to the device and technology, the latter refers to the behavior of the mobile worker that changes with context. From there, a strategy to address KLR through mobile devices should be built on an understanding of mobility (Agudelo et al., 2016; Derballa and Pousttchi, 2006a; Jiang et al., 2016)

| Context | Reference | Description |
|---|---|---|
| Environmental | Kofod-Petersen & Cassens (2006); Nieto, Botía, & Gómez-Skarmeta, (2006) | The environmental context is defined as the conjunction of the following contexts: temporal context, spatial context, social context, technological context, and business context |
| Personal | Kofod-Petersen & Cassens (2006) | The personal context provides the attributes of cognitive skills and draws on psychological and physiological contexts: psychological, goal, cognition, physiological, identity, actions |
| Social | Nieto et al., (2006) | Provides a social perspective of context, which captures the attributes of people (e.g. attitude, skills, and values) and the relationship of these people among each other and within the organization and collective structures (collective values and norms). |
| Technological | Abdoul Aziz Diallo (2012) | Provides the technological and technical attributes such as: network connections, infrastructure, equipment, devices and systems. It is an aggregate context which consists of other technical constituents such as spatial, user and location context. |
| Organizational | (Crossler et al., 2013; Furnell and Rajendran, 2012; Whitman, Michael and Mattord, 2011) | Defines the social interactions within the workplace and security behavior determined by Information Security Policies, Security Education Training and Awareness, Culture, Standards, organizational processes and procedures |
| Device | (Diallo et al., 2014; Kofod-Petersen and Cassens, 2006; Nieto et al., 2006) | Technological features such as device identifier, device type and processing capabilities (i.e., laptop, tablet, smartphone) |

*Table 1. Taxonomy of Relevant Mobile Usage Contexts derived from the literature adapted from (Agudelo-Serna et al., 2017)*

Mohamed et al. (2006) found that one of the key routes of knowledge leakage is *people* through social contexts of mobile usage. These routes include training courses, collaborations with universities, multi-disciplinary teams and temporary workers. Through social interactions in these different contexts, knowledge is shared or accessible to other users. Social context also includes the use of social networking platforms on mobile devices (Krishnamurthy and Wills, 2010).

Due to the nature of mobile device usage, the context of a device usage transitions across many changes in technical, social and locational environments (technological, environmental, organizational and personal contexts). Through the interactions of these dynamic contexts with one another, the risk of knowledge leakage also becomes dynamic. Thus, knowledge can be leaked through the technological, organizational, personal, and social contexts, amongst others (Diallo et al., 2011, 2014). As an illustration of this phenomenon, Astani et al (2013) found that a significant amount of employees from information-sensitive industries such as banking, connected their mobile devices to unsecured public Wi-Fi networks (i.e., technological context, environmental context) which exposes the device to the security vulnerabilities of those networks and may be used as a vehicle for knowledge leakage. By simply changing the network connection to a public Wi-Fi network, these employees are drastically changing the technological and environmental contexts and, therefore, their "mobile device usage context" in which the device is operating, changing the risk profile of their device, drastically affecting the potential for knowledge leakage.

These contexts are relevant to the use of the mobile device. If a user changes devices (device context, technological context), for example, then his/her overall context (user context) will change. The new device may not have the same functions as the previous one, resulting in a new number of contexts affecting the device. Since the old device is no longer used by the user, various contexts (e.g., social, user, and location contexts) no longer apply to it. This highlights the dynamic changes in knowledge leakage risks as the circumstances of how the knowledge worker uses their mobile device change.

Additionally, people and objects are constantly moving in and out of different context risks and the relevance of these objects and people to the context are dynamic and hence the security threat of knowledge leaking is constantly changing. For example, if Mary is sitting in a coffee store reading corporate emails from a tablet before heading into work (environmental, personal and technological context) and a new customer sits down behind Mary (social context), Mary's risk context has changed as the customer may potentially read Mary's tablet screen (shoulder surfing). Mary then receives a phone call (personal and social context), which introduces a new person (caller) into the context, with whom the agenda of a morning meeting is discussed (organizational context). This change in context risk now involves the surrounding people within earshot drastically increasing the potential for knowledge leakage.

From the literature there have been many approaches to modelling the contextual information surrounding mobiles across many disciplines of Information Technology. Most of the research into the *contextual information* and *context* of mobile devices has been focused on the technical and computing issues (Benítez-Guerrero et al., 2012; Bradley and Dunlop, 2005; Diallo et al., 2013; Hofer et al., 2003; Kofod-Petersen and Cassens, 2006; Schilit et al., 1994).

Similarly, Hofer et al. (2003) also extended and modelled these dimensions of context into device context (e.g. device identifier and device type) and network context (e.g. network connection types) which were included as the technical context, in a more recent study, by Abdoul Aziz Diallo (2012). However, these studies failed to address the social context, neglecting the human perspective from the mobile contextual model, namely, user behavior.

Conversely, Chen & Nath (2008) asserted that the social context is not independent of the technical context; it is the "interaction and compatibility" between the two that determine the effectiveness of a working system. This interdependency of the social and technical context is further reflected by Bradley and Dunlop's, (2005) "Model of Context in Computer Science" which aims to illustrate the key components and characteristics of context which are present during user-computer interaction. The key idea derived from Chen and Nath's model of context is that there are multiple contexts that contribute to the mobile usage context of mobile devices.

Expanding on Chen & Nath's (2008) social context interaction framework and Bradley and Dunlop's (2005) model of context, we address the gap in the literature by modelling such contexts from the human perspective and defining a high-level construct, knowledge leakage risk through mobile devices, as a formative construct (i.e., comprised of mobile usage contexts) which in turn informs risk mitigation strategies in organizations. This conceptual model is further explained in the following section.

## 3    Research Model

Figure 1 depicts our proposed research conceptual model. We develop our research model by identifying key constructs based on the two models mentioned in the previous section: Chen and Nath's (2008) "social context interaction framework" and Bradley and Dunlop's (2005) "model of context in computer science".

The criteria to select contexts for the conceptual model were based on the social context interaction framework (Chen and Nath, 2008). 1) Personal context and 2) Social context are grouped together under Human factors which refer to motivations and cognitive processes, as well as social norms that are explicit and implicit from human behaviors and social interaction. 3) Environmental context and 4) Organizational context constitute the Enterprise factors and refer to the organizational culture and behavior, operating environment (regulations) not only within the workplace but also outside (macro environment). Finally, the Technical factors are composed of 5) Device context and 6) Technological context and refer to the technology and information systems that enable and facilitate the adoption of technology and technical artefacts to perform knowledge-sharing activities.

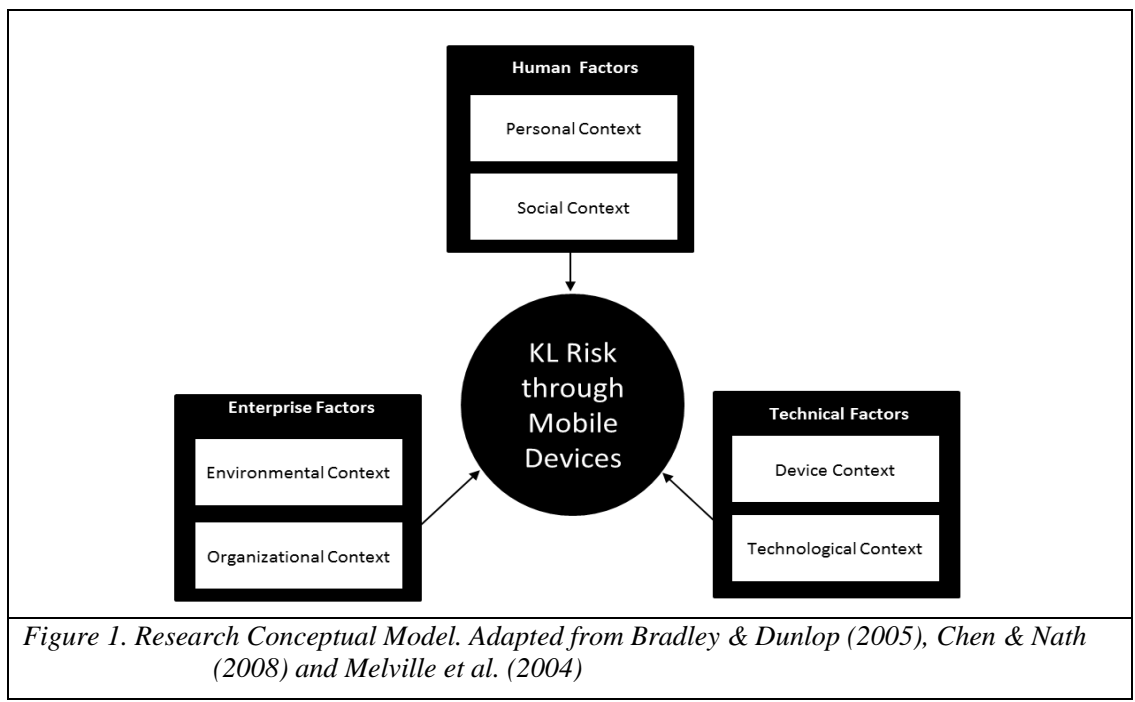

*Figure 1. Research Conceptual Model. Adapted from Bradley & Dunlop (2005), Chen & Nath (2008) and Melville et al. (2004)*

The constructs have been clustered in three groups: human, organizational and technical factors as defined in the "Integrative Model of IT Business Value" based on the resource-based view of the firm (Melville et al., 2004) as this model provides a framework to understand how internal (organizational resources) and external (trading partners, competitive and macro environment) factors impact organizational performance, which in our case, relates to how the mobile device usage contexts through the construct (i.e., KLR) contributes to improvement of organizational information and knowledge security performance signified in the organizational KLR mitigation strategies. In Table 2 the propositions are listed and explained with references to the literature.

## 3.1    Knowledge Leakage Risk caused by the use of mobile devices

The resource-based view (RBV) is used to determine the strategic resources in organizations with the potential to deliver comparative advantage to a firm. These resources can be exploited by the firm in order to achieve sustainable competitive advantage. RBV proposes that firms are heterogeneous because they possess heterogeneous resources, meaning firms can have different strategies because they have different resource mixes. Moreover, the RBV focuses managerial attention on the firm's internal resources in an effort to identify those assets, capabilities and competencies with the potential to deliver superior competitive advantages (Barney, 1991; Leonard-barton, 1992; Wernerfelt, 1984).

RBV also highlights the importance of protecting resources and capabilities to sustain competitive advantage in organizations (Leonard-barton, 1992). In this regard, the organizational knowledge capability needs to be protected, and the associated knowledge leakage risk (KLR) must be assessed. However, the risk evaluation process is subjective in nature which leads to a perceived KLR characterized by the impact and likelihood of leakage happening (ISO/IEC 27005:2011 2011). Consequently, the risk treatment involves selecting one or more options for modifying either the risk impact, probability or both. Such treatment includes implementing controls and strategies to treat the residual risks that are suited to the risk profile of the organization, environment and resources. Moreover, mobile workers and mobile devices further exacerbate the risk of leakage, and as a result, it becomes paramount to address the factors that such mobility brings.

| Construct | Definition | Reference |
|---|---|---|
| Knowledge leakage Risk through Mobile Devices | Knowledge leakage risk caused by the use of mobile devices in organizations. This high-level construct will be operationalized used a qualitative scale, i.e., low, medium and high. | (27005:2011, 2011; Agudelo et al., 2015; Ahmad et al., 2015) |
| Human Factors | The combination of personal and social contexts referring to individual's behavior, attitude, cognitive capabilities, motivations, experiences (personal context) as well as group's culture and values, social norms, peer's influence and superior's influence (social context). | (Ajzen, 1991; Bandura, 1978; Bradley and Dunlop, 2005; Chen and Nath, 2008; Melville et al., 2004) |
| Enterprise Factors | The combination of environmental and organizational contexts referring to external conditions (e.g., competitors, industry, external locations) as well as internal organizational resources and capabilities (e.g., policies, culture, processes, routines). | (Ajzen, 1991; Bradley and Dunlop, 2005; Chen and Nath, 2008; Melville et al., 2004) |
| Technical Factors | The combination of device and technological contexts referring to the infrastructure and technological resources internal and external to the organization that enable and support knowledge-sharing activities. | (Ajzen, 1991; Bradley and Dunlop, 2005; Chen and Nath, 2008; Melville et al., 2004) |

*Table 2. Definition of constructs in the research conceptual model*

Thus, drawing on RBV and the contextual framework, previous studies have evaluated a considerable number of organizational characteristics as determinants of competitive advantage, which in turn have been classified within the broader category of basic competences or influencing factors (Chen and Nath, 2008; Leonard-barton, 1992). In this research, we draw on these elements and extend them to include *human, enterprise and technical factors which affect the likelihood/impact of KLR in the context of mobile devices.* This research model characterizes the distinctive but complementary contexts within which leakage occurs that determines the risk exposure.

# 4      Methodology

Given the explorative nature of this study, we followed a qualitative research design using different participants. Data collection comprised 19 interviews (see appendix A for interview guide) of information security and knowledge managers of medium to large knowledge-intensive organizations in Australia (see table 3).

We selected knowledge-intensive organizations that already had mobile policies in place and allowed BYOD (Bring your Own Device) programs for their staff. Further, organizations sampled for this study fulfil the definition of *Knowledge-Intensive* organizations in that they were strictly associated with the cognitive assets they possess and their main production factor and outcome consisted of knowledge, directly delivered to customers in the form of consulting, or embedded in artefacts and services. Additionally, their activity mostly based on the exploitation of the workers' knowledge, specializations and skills (Bolisani et al., 2013; Tseng et al., 2011). Examples of knowledge-intensive organizations include software, telecommunication, pharmaceutical and consultancy companies.

Supplementary documentation (policies, procedures, reports and organizational standards) provided by the organizations was examined for triangulation as a way to confirm that the data reported in interviews matched organizational documented processes and procedures. To analyze the supplementary data, a content analysis on documentation was conducted looking for supporting evidence on the secondary sources that corroborated the interviews statements. In many cases, information provided by interviewees did not match the organizational documents, highlighting the issue of either outdated or incomplete documentation. In other cases, the practices reported by participants were not formally institutionalized but rather a cultural practice (informal mechanisms). By conducting an organizational document review, we were able to establish what strategies are formally conducted by organizations and what strategies, although informal, are rooted in the organization's culture.

| ID | Role | Industry | Knowledge Asset(s) | Experience (Years) | Self-Reported Risk of Leakage |
|---|---|---|---|---|---|
| CIO1 | Chief Information Officer | Government | Policy, Processes, Strategy | 10+ | Medium |
| SM1 | Security Manager | Banking | Strategy, Processes, | 15+ | Medium |
| CISO 1 | Chief Information Security Officer | Consultancy | Processes, Product, Methodology | 20+ | High |
| SM2 | Security Manager | IT Provider | Product, Methodology | 10+ | High |
| CTO1 | Chief Technical Officer | IT Services | Intellectual Property, Product, Methodology | 15+ | High |
| SM3 | Security Manager | Insurance | Processes, Methodology | 10+ | High |
| CISO 1 | Chief Information Security Officer | Health Care | Product | 10+ | Medium |
| CSM | Cyber Security Manager | Consultancy | Processes, Methodology | 15+ | High |
| CISO 2 | Chief Information Security Officer | Telecommunications | Product, Methodology | 15+ | High |
| CI-KO[1] | Chief Information and Knowledge Officer | Government | Policy, Strategies | 10+ | Medium |
| CKO1 | Chief Knowledge Officer | Food | Intellectual Property, Process, Product | 15+ | High |
| KM1 | Knowledge Manager | Health Care | Client, Product, Processes | 10+ | Medium |
| KM2 | Knowledge Manager | Government | Policy, Strategy | 10+ | Medium |
| CKO2 | Chief Knowledge Officer | Consultancy | Product, Methodology | 15+ | High |
| KM3 | Knowledge Manager | Consultancy | Processes, Methodology | 10+ | High |
| CKO3 | Chief Knowledge Officer | Government | Processes, Strategy | 15+ | Medium |
| KM4 | Knowledge Manager | Health Care | Product , Processes | 5+ | Medium |
| KM5 | Knowledge Manager | Education | Methodology, Research findings | 10+ | Medium |
| KM | Knowledge Manager | Not-for-Profit | Client relationship | 15+ | Low |

*Table 3. Information Security and Knowledge Managers participants*

The purpose of each interview was to establish how each organization encouraged the flow and sharing of knowledge, particularly through mobile devices, while also ensuring that knowledge leakage does not occur. The reason we interviewed senior managers was to identify what strategies they had in place to prevent knowledge leakage. Interviews were conducted over a period of 9 months. Each interview lasted approximately 1 hour and was audio-recorded with the consent of each interviewee and transcribed verbatim and shared with each interviewee to check validity and verify the content. During the interviews different context scenarios were presented to illustrate different levels of risk exposure (See Appendix B for examples of scenarios provided to participants). The scenarios illustrated the relationship between contexts and risk. Additionally, as previously indicated, documents such as policies and procedures were also analyzed to get a better understanding of knowledge protection mechanisms used in the organizations. The transcribed data was analyzed using selective, axial and thematic content analysis (Krippendorff 1980; Miles & Huberman 1994) and drawing on the different mobile contexts outlined in the research model to classify collected evidence (see figure 2). The Findings in terms of the different knowledge protection controls and mechanisms used by security and knowledge professionals in knowledge-intensive organizations are summarized in table 4.

---

[1] *This participant was considered in both groups due to his expertise in both fields.*

# 5    Findings

The findings in this section list the leakage strategies reported by security and knowledge managers of knowledge-intensive organizations in Australia that were part of this study. Given that our sample was specific and relatively small, it is not generalizable. However, it is indicative of the practices followed by many organizations in Australia.

In summary, the findings of this study show that many of the strategies overlap the human, enterprise and technical factors. Notwithstanding, security and knowledge managers in knowledge-intensive organizations in Australia are aware of the differences, the focus is now shifting from the enterprise and technical dimension to the human factors. Organizations are increasingly aware of the current threat landscape, and the only way to keep current and maintain competitive advantage in an environment that is ever-changing and increasingly complex is through protection of resources and organizational capabilities, such as knowledge and information assets.

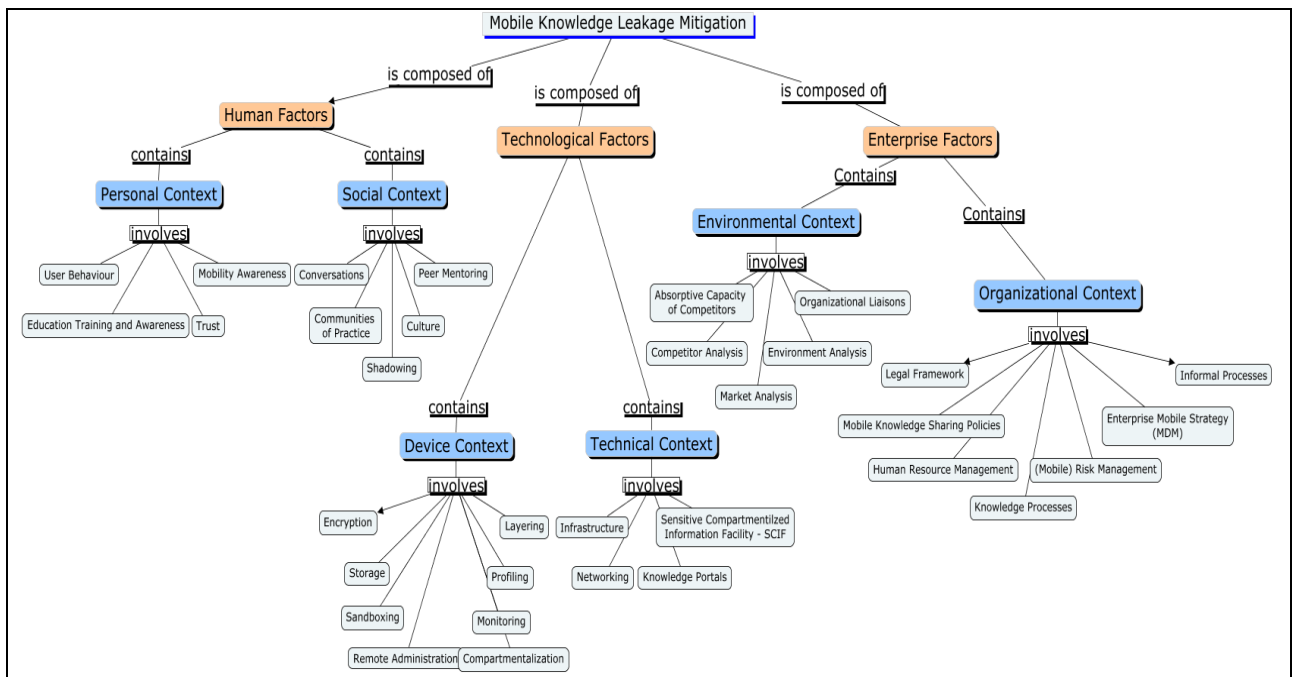

*Figure 2. Strategies used by knowledge intensive organizations*

Similarly, strategies that focus on employees' behaviors as well as behavioral change are just as important. The key is to increase employees' understanding and awareness that the way they interact with other people, their mobile devices and computing systems can enhance or diminish the effectiveness of a security program. Sustained periodic training and awareness also serve to reinforce policies and procedures in the minds of employees.

By analyzing the data collected from the interviews using content analysis and drawing on the research model, we have provided a framework to classify the evidence and a tool for practitioners that can be used as a guide and checklist to combat knowledge and information leakage caused by the use of mobile devices in knowledge-intensive organizations in Australia.

| Factor | Context | Organizational Strategy | Source of Evidence | Quote |
|---|---|---|---|---|
| **Human** | Personal | -User Behavior Analytics (Artificial Intelligence)<br><br>-Gamification (Simulation of scenarios)<br><br>- Education, Training and Awareness for knowledge sharing through mobile devices | Policy, Interview | *"our CEO always says: if people are the weakest link then education is the strongest link, this is why we are investing in a comprehensive education program for our staff" [SM2]* |
| | Social | -Develop a Security Culture<br><br>-Knowledge Communities<br><br>-Peer Mentoring Policy | Policy Interview | *"The good thing about the program that we started a few years ago is that now the employees teach the new staff our values and principles and often I see how they report any suspicious activity before actioning on requests coming from strangers or emails " [KM4]* |
| **Enterprise** | Organizational | -Risk Management Framework (Identification of valuable knowledge assets)-<br><br>-Knowledge Management Strategy<br><br>-Develop resilience capability<br><br>-Develop policy/procedures/guidelines on knowledge sharing through mobile devices<br><br>-Knowledge reconfiguration (combine knowledge assets to create new knowledge)<br><br>-Legal Frameworks (Non-disclosure agreements, contracts, patents)<br><br>-Multi-disciplinary integration among org. areas (e.g., HR, IT, Legal, Finance) | Policy Interview | *"We have a really strict screening policy, once a knowledgeable person leaves the organization, in fact, the policy states that screening is on-going. So when you join the company you have to undertake a long screening process and after that every year HR reminds us the process and even when you leave you need to follow an exit policy to make sure there is no liability for the company " [KM1]* |
| | Environment | - Liaisons between organizations ( Government, Research, private and public sector)<br><br>-Market analysis (Fast Innovation Cycle)<br><br>-Competitor/Adversary Analysis (Tactics, Motivation, Goals) | Interview | *"So far we have different partnerships with universities in the UK and the US to help us with research and development of technologies, however we only give them the bare minimum just to make sure there's no chance of a breach and they usually work in another location isolated from us " [CISO2]* |
| **Technical** | Device | -Mobile Device Management (MDM)<br><br>-Geolocation settings<br><br>-Device Sandboxing<br><br>-Remote administration<br><br>-Device Profiling (Usage Pattern) | Policy Procedure | *"We use a feature within Airwatch that is called Secure Content Management that allows our mobile force to access documents on the go through their laptops or iPads but the physical location of the document is on our servers so if anything happens we just revoke access to the content without messing with their actual equipment " [SM3]* |
| | Technological | -Enterprise Mobility Strategy<br><br>- Mobile Device Usage Analytics (Artificial Intelligence)<br><br>-Knowledge Compartmentalization (access, clearance)<br><br>-Knowledge Classification (tagging, labelling) | Policy Procedure | *"What I love about our SIEM is that it displays a nice dashboard showing the patterns for a particular individual, so we know more about their usage and their profiles and sometimes it even notifies us when a possible person may be at risk of leaving the organization as their behaviour changes and, for instance, starts sending a lot of company information to other accounts outside our authorized domains.[SM1]* |

*Table 4. Summary of knowledge leakage strategies observed in the study.*

## 5.1 Proposed Knowledge Leakage Mitigation Framework for Mobile Devices

Based on the interviews and the conceptual model, we categorized the reported mitigation strategies drawing on two dimensions grounded in literature, that is, degree of formality of leakage mitigation strategies – formal vs informal - (Amara et al., 2008; Bolisani et al., 2013; Dhillon, 2007; Sveen et al., 2009) and type of knowledge –Tacit vs Explicit - (Nonaka, 1991, 1994). The framework is depicted in figure 3, the first quadrant, 'Formal Tacit', applies to mature organizations that have processes in place to protect knowledge that has not been articulated yet. Examples of typical organizations in this quadrant are intelligence and military organizations. The second quadrant 'Formal Explicit' best suits organizations that need to protect codified knowledge from competitors, for example, software and service companies. The 'Informal Tacit' quadrant is well suited for highly-innovative organizations that need to protect intellectual property but do not have knowledge processes in place (i.e., small organizations, start-ups). The last quadrant 'Informal Explicit' is usually utilized by organizations that rely on codified knowledge but do not have proper secure knowledge processes in place. It is worthwhile noting that, these strategies are not mutually exclusive; rather they can be used in conjunction.

Our findings suggest that informal protection methods are more adopted than formal methods. Formal methods are based on legal measures and organizational processes. These methods comprise policies and procedures that manage the access to and use of knowledge. Formal methods include policies and procedures that establish and ensure the effective use of technical controls. Examples of such mechanisms include system audits, update mechanisms, risk assessments, identification of security roles, segregation of responsibilities and implementation of indicators. In contrast, informal protection methods are based on relationship, trust and organizational arrangements. Typically, these methods involve actions related to deploying security in organizations by creating a security culture. Examples of informal controls include training employees, implementing security incentives, increasing the commitment to security and motivating users.

Among informal methods, some are more popular (for example, secrecy, trust, and fast innovation cycles). Organizations tend to use different types of strategy (formal, informal) for achieving a better overall protection.

| Type of Knowledge | | |
|---|---|---|
| | **Tacit** | **Explicit** |
| **Formal** | **Human**<br>1. Peer mentoring<br>2. Education Training Awareness<br>3. Security Culture Development<br>4. Communities of Practice<br>5. Conversation Security (Phone)<br><br>**Enterprise**<br>6. HR Management<br>7. Knowledge Sharing Policy<br>8. Agreements with other organizations<br><br>**Technological**<br>9. Expert Directories<br>10. Case conferences<br>11. Content/ People separation | **Human**<br>1. Security Gamification<br>2. Mobile Behaviour Profiling/Analysis<br>3. On-going Screening / monitoring<br><br>**Enterprise**<br>1. Enterprise Mobile Strategy<br>2. Mobile Risk Management<br>3. Legal Framework (Contracts)<br>4. Industrial property Rights<br><br>**Technological**<br>1. Securing Communication channels and Devices<br>2. Compartmentalization<br>3. Device Monitoring / Detection<br>4. Perimeter Defence |
| **Informal** | **Human**<br>1. Trust development<br>2. Secrecy<br>3. Staff rotation<br>4. Staff Shadowing<br><br>**Enterprise**<br>1. Fast Innovation Cycle<br>2. Complexity in Knowledge processes<br>3. Lead Time Advantage<br><br>**Technological**<br>1. Deception (Decoy campaigns) | **Human**<br>1. Discretionary Access to knowledge-base<br><br>**Enterprise**<br>1. Knowledge-sharing with business partners<br><br>**Technological**<br>1. Restricted access |

*(Left vertical axis label: Knowledge Leakage Mitigation Strategy Type)*

*Figure3.  Knowledge Leakage Mitigation framework for Mobile devices based on interviews*

# 6        Discussion and Conclusion

The results of this empirical study are expected to have both practical and theoretical implications. This study is expected to contribute to IS security research by proposing a comprehensive conceptual model that will be empirically tested in later phases and will investigate the determinants of knowledge leakage risk through mobile devices in knowledge-intensive organizations operating in highly competitive environments in Australia. Our study is also expected to provide meaningful implications for security and knowledge managers in organizations to improve risk mitigation strategies, policies and education and training associated with knowledge leakage.

In today's security landscape, mobile devices pose new threats to organizations' security and knowledge management strategies. Effective KLR mitigation strategies will help organizations better manage those devices in their environment protecting their organizational knowledge. This study is the first attempt to view KLR through mobile devices in organizations from a mobile usage perspective using a contextual approach combining human, enterprise and technological dimensions. By analyzing the determinants that influence the knowledge leakage risk through mobile devices in organizations, addressing not only technological aspects but also human and organizational aspects, the proposed model presents a better way to design mitigation strategies and leakage risk controls (i.e., formal, informal and technological) that are more likely to be accepted and followed by employees.

We have presented a conceptual framework that seeks to explain how the knowledge leakage risk is influenced by human, enterprise and technological factors and how such KLR informs organizations' mitigation strategies. Empirical confirmation and refinement of the research conceptual model is an important future research direction to follow. Additionally, we have categorized the findings of the interviews on leakage strategies used by knowledge-intensive organizations in Australia drawing on the conceptual model proposed.

However, our study has a number of limitations. First, our sample was specific and relatively small. The findings in this study need to be explored in larger empirical studies across multiple organizational sectors. Second, our main source of information was interviews with senior-level managers. As such, we did not explore leakage-related behaviors at the operational level in terms of operational staff, which, in turn, points to the need for further studies in terms of actual employee behavior.

This study is the first part of a larger study. In this current phase, we surveyed security and knowledge managers about the different strategies and mechanisms used to address knowledge leakage caused by mobile devices in different contexts. In this phase, 19 interviews with knowledge managers (9) and security experts (10) were conducted. The objective of this phase was to better conceptualize the different constructs and investigate how such factors characterize KLR mitigation strategies using the conceptual model as reference.

In a future second phase, we will conduct two focus groups, one with knowledge managers and another with security managers from different knowledge-intensive sectors in Australia to further improve the concepts and the underlying propositions in the model. The goal of this phase is to develop specific-sector insights (i.e., private, military, governmental and not-for-profit organizations) in order to contrast different industries.

In a third phase, we will conduct in-depth multiple case studies, following Yin' s (2003) methodology, in different types of knowledge-intensive organizations (e.g., private, military, governmental and not-for-profit organizations) which operate in highly-competitive environments in order to validate our findings and further refine the proposed conceptual model from our previous phase. The objective of this phase is to generalize the findings of this research to other industries.

# Appendix A – Interview guide

**Background questions:**
1. What is your current position at your organization, years of employment, experience, academic training, and industry?
2. In your organization, is knowledge sharing critical for the work of your employees?

**Opening questions**
1. What is your general perception of knowledge leakage issues?
2. Have you experienced knowledge leakage issues in you organization?
3. How often does the topic of knowledge leakage arise? In what situations does it happen?

**Scenario Questions: Considering the scenarios provided to you, please answer the following questions:**
1. Have any of the scenarios provided to you resembled a situation in your organization?
2. Do you have policies in place that address this behaviour in your organization? Please explain.

**Given the distinction between knowledge and information [Definitions and distinctions given to participants in advanced]:**
1. What current mechanisms do you use to protect *information?*
2. What current mechanisms do you use to protect *Knowledge?*
3. What are the main knowledge assets you organization need to protect?
4. Do you have a risk management procedure or strategy in your organization? If yes, how is knowledge managed?
5. What knowledges processes does your organization have in place? (Examples: Capture, retention, transfer, storage)
6. In your opinion, what knowledge is the most critical and therefore should be protected? (Example: Intellectual Property, process knowledge, marketing strategies, client knowledge, etc.)
7. Do you think your knowledge assets are at risk of being appropriated by partners/competitors/clients? Please explain.
8. What would happen if these knowledge assets leaked? What's the potential impact? How likely is it to happen?
9. Based on your experience, how does knowledge leakage occur?
10. How do you discourage/encourage knowledge sharing in your organization?
11. How do you currently share sensitive knowledge among employees, partners and clients?
12. What risk assessments do you undertake when sharing knowledge among employees, partners and clients?
13. Do you protect your knowledge assets in cooperation/transactions/ partnerships with partners/competitors/clients? Please explain.
14. What controls do you have in place to protect these assets?
15. Do you think the use of mobile devices such as smartphones, tablets, and laptops are beneficial to knowledge creation/sharing? Please explain.
16. Do you think the use of mobile devices such as smartphones, tablets, and laptops pose a greater risk? Please explain.
17. Do you have an organizational mobile policy/Strategy? Please explain.
18. How do you secure the mobile devices in your organization when used in knowledge sharing activities?

19. Does your firm have any formal or informal routines for dealing with potential knowledge leakage through mobile devices? Please explain.
20. Based on your experience, what protective actions do you think should be used to address the risk of knowledge leakage caused by the use of mobile devices?

## Appendix B – Example of Scenarios provided to participants

| Scenario 1 |
| --- |

Stephen, a senior manager, is in an airport waiting for his delayed flight to a business conference. The conference will be attended by many senior professionals within the industry. The following events transpire in the day:

    A) While waiting in the crowded airport lounge, Stephen connects his Wi-Fi only tablet to the airport public Wi-Fi. There were multiple public Wi-Fi so Stephen chose the one with the strongest signal

    B) Stephen receives an email from John regarding a tender proposal and downloads it onto his tablet to review

    C) The airplane has finally arrived and Stephen proceeds to the waiting area which had a different network which he used while waiting to board.

    D) After finding his seat in the middle of the aisle, Stephen noticed that there was in-flight Wi-Fi access so he decides to log into it.

    E) After reviewing and commenting on the tender proposal, Stephen sends the revision to John via corporate email

Stephen checks into his hotel room and reads tomorrow's conference agenda on the hotel's Wi-Fi

**Do any of Stephen's actions relate to you and your working experiences/habits? (Please indicate below by circling the corresponding letters)**
A  B  C    D    E    F

**Did Stephen's actions break any of your company's policies in this scenario? (If yes, please indicate which actions below by circling the corresponding letters)**
A  B  C    D    E    F

**Do you think any of Stephen's working habits present a security risk? (If yes, please indicate below by circling the corresponding letters)**
A  B  C    D    E    F

| Scenario 2 |
| --- |

Joseph is in the workplace, finishing off a large tender proposal. Work is unusually hectic on a Thursday afternoon as the deadline to submit the tender is Friday. The following events transpire:

    A) Joseph is at work, using a laptop connected to the company Wi-Fi to finish the report

    B) When work ends, Joseph takes his laptop home. The train ride home was very crowded Joseph was able to get a seat and review documentation on his tablet

    C) During the train ride, Joseph receives an urgent call from his boss, asking Joseph to finish the report as soon as possible.

    D) Joseph takes out his laptop and starts making changes to the report requested by his boss before he forgets. The passenger next to Joseph moves over to give him extra room to use his laptop and tablet.

    E) After arriving home, Joseph makes the final changes to the tender proposal and emails it to his boss.

    F) To relieve stress from the week, Joseph goes to his local bar. As the bar is underground with poor network reception, Joseph's phone automatically connects to the bar's Wi-Fi

    G) Joseph's boss responds to his email by resending the tender proposal document with added comments for changes. After reading the comments, Joseph finishes his drink and heads home to continue working.

**Do any of Joseph's actions relate to you and your working experiences/habits? Please indicate below by circling the corresponding letters**
A  B  C    D    E    F    G

**Did Joseph break any of your company's policies in this scenario? If yes, please indicate below by**

| circling the corresponding letters: |
| --- |
| A    B    C         D         E         F         G |
| **Do you think any of Joseph's working habits present a security risk? (If yes, please indicate below by circling the corresponding letters)** |
| A    B    C         D         E         F         G |

**Scenario 3**

Stephanie is working casually for a small company, while completing her Master's degree part-time, and was recently allocated a tablet device. On a typical day:

- A) Stephanie uses her tablet to take pictures of whiteboards and presentation slides during meetings, as well as audio recordings
- B) Uses her tablet to access personal email accounts and browse her favourite websites, such as Facebook, during breaks on the organisation's Wi-Fi
- C) Download applications onto her tablet for work as well as entertainment and social networking
- D) Review reports and documentation
- E) Access the wiki knowledge management portal to upload documentation and update project progress information

Stephanie also takes the tablet to University to write notes during lectures, complete assignments and study for tests and exams

| **Do any of Stephanie's actions relate to your working habits? Please indicate below by circling the corresponding letters:** |
| --- |
| A    B    C         D         E         F |
| **Did Stephanie break any of your company's policies in this scenario? If yes, please indicate which of her actions by circling the corresponding letters:** |
| A    B    C         D         E         F         G |
| **Do you think any of Stephanie's working habits present a security risk? If yes, please indicate below by circling the corresponding letters:** |
| A    B    C         D         E         F         G |